# One MAX phase, different MXenes: a guideline to understand the crucial role of etching conditions on Ti$_3$C$_2$T$_x$ surface chemistry

Mohamed Benchakar,[a] Lola Loupias,[a] Cyril Garnero,[a,b] Thomas Bilyk,[b] Cláudia Morais,[a] Christine Canaff,[a] Nadia Guignard,[a] Sophie Morisset,[a] Hanna Pazniak,[b] Simon Hurand,[b] Patrick Chartier,[b] Jérôme Pacaud,[b] Vincent Mauchamp,[b] Michel W. Barsoum,[c] Aurélien Habrioux,[a,*] Stéphane Célérier [a,*]

[a] Institut de Chimie des Milieux et Matériaux de Poitiers (IC2MP), Université de Poitiers, CNRS, F-86073 Poitiers, France

[b] Institut Pprime, UPR 3346 CNRS, Université de Poitiers, ISAE-ENSMA, BP 30179, 86962 Futuroscope-Chasseneuil Cedex, France

[c] Department of Materials Science and Engineering, Drexel University, Philadelphia, PA 19140, United States

**Corresponding Authors:**

* E-mail: aurelien.habrioux@univ-poitiers.fr, stephane.celerier@univ-poitiers.fr

ABSTRACT

MXenes are a new, and growing, family of 2D materials with very promising properties for a wide variety of applications. Obtained from the etching of MAX phases, numerous properties can be targeted thanks to the chemical richness of the precursors. Herein, we highlight how etching agents govern surface chemistries of $Ti_3C_2T_x$, the most widely studied MXene to date. By combining characterization tools such as X-ray diffraction, X-ray photoelectron, Raman and electron energy loss spectroscopies, scanning and transmission electron microscopies and a surface sensitive electrochemical reaction – the hydrogen evolution reaction, HER – we clearly demonstrate that the etching agent (HF, LiF/HCl or $FeF_3$/HCl) strongly modifies the nature of surface terminal groups (F, OH and/or O), oxidation sensitivity, delamination ability, nature of the inserted species, interstratification, concentration of defects and size of flakes. Beyond showing how using these different characterization tools to analyze MXenes, this work highlights that the MXene synthesis routes can influence targeted applications.

1. Introduction

MXenes are among the newest and largest family of 2D materials with demonstrated applications in diverse fields such as electromagnetic interference shielding [1], transparent conductive films [2], wireless communication [3], hydrogen and oxygen electrocatalysis [4–7], supercapacitors [8], batteries [8], catalysis [9], photocatalysis [10], hydrogen storage [11], gas and biosensors [12,13], electromechanical actuators [14], pH and humidity sensor [15,16], nanomedicine [17] or gas and liquid separation and purification[18,19]. The properties and potential applications of these materials have been the subject of several recent reviews[20–25].

This very broad spectrum of applications is mostly due to the high versatility of the MXene composition and thus of their related properties. Indeed, these materials are synthesized by selectively etching the A element from the $M_{n+1}AX_n$ phases, a family of about 150 different members of layered ternary carbides, nitrides or carbonitrides [26], where M is an early transition metal, A is an element mostly from columns 13 or 14 of the periodic table (*i.e.* group IIIA or IVA), X is C and/or N and n = 1 to 3, leading to the formation of many different $M_{n+1}X_n$ metal carbides and/or nitrides [27]. To date, more than 30 different MXenes have already been synthesized and many others have been theoretically predicted [24],[28]. Beyond the possibility to tune the MXene composition and related properties by changing the MAX phase precursor, it is well-established that different etching protocols lead to $M_{n+1}X_nT_x$ flakes with different T terminal groups (F, OH and/or O), that can also strongly modify MXene properties [29–32]. Although crucial for many applications, the effects of MXenes functionalization on properties is still in its infancy because of the limited number of etching processes and the need to design characterization protocols that can accurately quantify the nature of these terminations and relate them back to the synthesis methods.

To date, several etching agents have been studied, the most common of which are HF [33], $NH_4HF_2$ [34], LiF in HCl or LiCl in HF [8,35]. Moreover, LiF can be replaced by other salts such as NaF, KF, CsF [8]. More recently transition metal-fluorides were used, such as $FeF_3$ and $CoF_2$ [36,37]. Non-fluorinated agents, such as alkali based compounds have also been used [38–40]. Other methods, including electrochemical etching [40,41] and etching with molten salts [42], have also been used. The choice of the etching agent has been shown to have a large impact on the surface chemistry, oxidation sensitivity, hydrophilic properties, delamination ability, conductivity, nature of inserted cations, size of flakes and degree of defects among other properties [36,43].

Although several studies compared the effect of different etching agents [36,43], a straight and systematic comparison of the $Ti_3C_2T_x$ surface chemistries obtained using the *same* initial MAX phase powder and combining characterization tools such as X-ray diffraction (XRD), X-ray photoelectron spectroscopy (XPS) and Raman spectroscopy in the *same* study is currently lacking in the literature. Additionally, many studies have focused on the etching of $Ti_3C_2T_x$ using HF but the etching conditions, such as temperature, duration, HF concentration can vary considerably from one report to another. This is mainly due to the fact that the structural and physicochemical properties of the initial $Ti_3AlC_2$ powders (*e.g.*, grain size, crystallinity, purity, synthesis protocol) vary, thus rendering a direct comparison between different studies difficult. For example, Kong *et al.*[44] showed that the $Ti_3AlC_2$ synthesis method (combustion or pressureless synthesis) significantly affects the electrochemical properties of the obtained MXenes. Shuck *et al.* [45] illustrated also the major role of the MAX synthesis method on the corresponding MXene properties. The use of the same initial MAX phase is therefore crucial for a rational study of the role of the etching agents on MXene properties.

The purpose of this work is to analyze the influence of different etching agents on the surface chemistries of $Ti_3C_2T_x$ multilayers obtained from the same MAX phase batch. We chose to focus on three different etching agents: HF, LiF/HCl and $FeF_3$/HCl as they result in very different surface chemistries. Moreover, for each etching agent, two different batches were prepared: for one batch, the etching conditions are described as soft; for the other the conditions were harsher. This work does not aim at precisely studying the effect of each synthesis parameter such as temperature or concentration for each etching agent, but it intends to highlight, for a given etchant, how the surface chemistries can be impacted by simply adjusting the synthesis conditions. Therefore, a complete set of characterizations (XRD, SEM, EDS, XPS, Raman, TEM-EELS) was performed to determine structural features of the different $Ti_3C_2T_x$ multi-layers. Obtaining significantly different surface states for $Ti_3C_2T_x$ was also used as a helpful database to confirm or rebut assignments of spectroscopic signals (Raman, XPS) proposed in the literature. This was also useful to provide a better understanding of differences observed in XRD patterns. Lastly, in order to further highlight the crucial role played by the synthesis conditions on the surface properties, the hydrogen evolution reaction, HER, was used as a probe reaction as it is mainly driven by the surface properties of the material under study.

## 2. Experimental

*2.1 Synthesis of materials*

To produce $Ti_3C_2T_x$ MXene powders (6 samples), 3 different etching agents were used: HF, LiF/HCl and $FeF_3$/HCl with two quite different conditions (labelled as soft and harsh). Systematically, 0.5 g of the $Ti_3AlC_2$ MAX phase precursor (initial particles sizes < 25 μm, see ESI part I for MAX synthesis) was used.

**$Ti_3C_2$-HF10 and $Ti_3C_2$-HF48.** The $Ti_3AlC_2$ powder was gradually added to 10 mL of hydrofluoric acid, HF, under stirring. [Safety Note: hazardous product requiring appropriate protocol]):

- 48 wt.% - Sigma-Aldrich - for $Ti_3C_2$-HF48 (**harsh conditions**)

- 10 wt.% for Ti$_3$C$_2$-HF10 (prepared by dilution of HF-48wt%) (**soft conditions**)

The mixture was stirred at 25 °C for 24 h before the centrifugation step.

**Ti$_3$C$_2$-Li-24/25 and Ti$_3$C$_2$-Li-72/60.** The Ti$_3$AlC$_2$ powder was gradually added to 10 mL of 0.8 g of LiF (≥ 99%, Sigma) dissolved in 10 mL of 9 mol L$^{-1}$ HCl (prepared by HCl 37%, Sigma Aldrich). Thus, the initial F/Al atomic ratio is of 12. The reaction was running for 24 h at 25 °C (Ti$_3$C$_2$-Li-24/25 – **soft conditions**) or 72 h at 60 °C (Ti$_3$C$_2$-Li-72/60 – **harsh conditions**) under stirring. This synthesis method is partly based on the work of Alhabeb *et al*. [43].

**Ti$_3$C$_2$-Fe-24/25 and Ti$_3$C$_2$-Fe-72/60.** Ti$_3$AlC$_2$ was gradually added to 10 mL of 1.716 g of FeF$_3$,3H$_2$O (Sigma-Aldrich) dissolved in 10 mL of 9 mol L$^{-1}$ HCl (prepared by HCl 37%, Sigma Aldrich). As in the case of LiF/HCl etching method, the F/Al atomic ratio is of 12. The reaction was running for 24 h at 25 °C (Ti$_3$C$_2$-Fe-24/25 – **soft conditions**) or 72 h at 60 °C (Ti$_3$C$_2$-Fe-72/60 – **harsh conditions**) under stirring. This synthesis method is partly based on the work of Wang *et al*. [36].

Except for Ti$_3$C$_2$-Li-24/25, the suspension obtained after the etching step was centrifuged 4 times at 6000 rpm for 6 min with ultrapure water (these washing cycles are sufficient to reach a supernatant pH value higher than 5) and the supernatant liquid was systematically removed. The resulting sediment was then filtered using a vacuum-assisted filtration device (PVDF membrane, 0.22 µm pore size), washed several times with ultrapure water, and dried one night at room temperature under air. As already proposed by some of us [36], the powders were finally washed in 350 mL of deaerated water during 4 h to remove aluminum impurities following by filtering and drying at room temperature under air during one night. The powders are then weighted and stored under inert atmosphere.

For Ti$_3$C$_2$-Li-24/25, the suspension obtained after etching, including Ti$_3$C$_2$T$_x$ and non-reacted Ti$_3$AlC$_2$, was centrifuged 8 times at 6000 rpm for 6 min and the supernatant liquid was systematically removed. The black Ti$_3$C$_2$T$_x$ slurry was then carefully collected and separated from the surface of the sediment Ti$_3$C$_2$T$_x$/Ti$_3$AlC$_2$ with a spatula as proposed by Alhabeb *et al.* [43]. The slurry and the sediment were then filtered separately, washed and dried one night under air at room temperature and finally weighted and stored under inert atmosphere.

*2.2 Characterization of materials*

XRD analysis of the MXene powders were carried out with a PANalytical EMPYREAN powder diffractometer using CuK$_\alpha$ radiation source (K$_{\alpha_1}$ = 1.5406 Å and K$_{\alpha_2}$ = 1.5444 Å). XRD patterns were collected between 5 and 70 ° with a 0.07 ° step and 420 s dwell time at each step. An ultra-fast X-Ray detector (X'Celerator) was used to collect the signals. The identification of Ti$_3$AlC$_2$, TiC, and LiF phases was performed with the HighScorePlus software (PANalytical©) and by comparison with the ICDD database reference files. Phase identification of MXene from the XRD patterns is based from our previous works [46].

The XPS analysis were carried out with a Kratos Axis Ultra DLD spectrometer using a monochromatic Al K$_\alpha$ source (1486 eV, 10 mA, 15 kV). When required, a charge neutraliser system was operated for the analysis. Instrument base pressure was 9×10$^{-8}$ Pa. High-resolution spectra were recorded using an analysis area of 300 μm×700 μm$^2$ and a 20 eV pass energy. These pass energies correspond to Ag 3d$_{5/2}$ FWHM of 0.55 eV. Data were acquired with 0.1 eV steps. All the binding energies were calibrated with the C1s (C-Ti-T$_x$) binding energy fixed at 281.9 eV as an internal reference since this component is present in all MXenes and it is very well distinguished from other surface contaminations contributions.

Raman spectroscopy was carried out using a HORIBA Jobin Yvon LabRAM HR800 confocal Raman microscope with a CCD detector. Spectra were acquired at RT using an excitation wavelength of 632.8 nm, supplied by an internal He-Ne laser. The power delivered at the sample is less than 1 mW. An 1800 grooves.mm$^{-1}$ grating is used resulting a spectral resolution of 0.5 cm$^{-1}$. The spectrometer is calibrated by a silicon wafer.

It is important to note here that all Raman and XPS spectroscopies were performed on fresh samples (i.e. less than 24 h after synthesis) except when it is specified (aged samples).

The Al, Ti, Li and Fe contents of the different $Ti_3C_2T_x$ were determined by Inductively Coupled Plasma-Optical Emission Spectrometry (ICP-OES) using a PerkinElmer Optima 2000DV instrument. The determined contents of each element were normalized to 3 Ti atoms.

The morphology of the MXene powders was studied using a field emission gun scanning electron microscope (FEG-SEM) 7900F from JEOL. This microscope is equipped with an Energy Dispersive X-ray Spectrometer (EDS) from Brücker (with Esprit software) allowing the determination of the O, F, Al, Ti, Cl and C contents. The determined contents of each element were normalized to 3 Ti atoms.

Specific surface area measurements were performed using nitrogen adsorption at -196°C with a TRISTAR 3000 gas adsorption system. Prior $N_2$ adsorption, the powder samples were degassed under secondary vacuum for 12 h at 80 °C. The BET equation was used to calculate the specific surface area of the samples ($S_{BET}$ in m$^2$ g$^{-1}$).

Transmission electron microscopy (TEM) experiments were performed on the $Ti_3C_2$-Li-72/60 sample. TEM samples were prepared by diluting the powder into water followed by ultrasonication for 2 minutes with a power of 150 W to separate the $Ti_3C_2T_x$ flakes. A drop of the colloidal suspension was then deposited on copper grid covered with a lacy carbon film. The observed areas on the sample were investigated using energy filtered electron diffraction (EFED) and electron energy loss spectroscopy (EELS). Both short range and long-range order was probed in the MXene structure. Experiments were performed in a JEOL 2200 FS TEM, operating at 200 kV, and equipped with an in-column omega filter. A 10 eV slit was used to acquire the EFED patterns. The energy resolution was about 1.0 eV and the collection angle was about 5 mrad. EEL spectra in the core losses were recorded. The backgrounds of the core-losses spectra were extracted using a power law and the multiple scattering effect was removed using the Fourier-ratio method.

*2.3 Electrochemical measurements*

All electrochemical measurements were carried out at RT in a standard three-electrode electrochemical cell using potentiostat Autolab (PGSTAT-302N) coupled with a rotating disc electrode (RDE) on fresh samples (unless otherwise stated). An Ag/AgCl (saturated KCl-filled) ($E_{AgCl/Ag}$ = 1.02 V *vs.* RHE) electrode and a glassy carbon plate were used, respectively, as reference and counter electrode. A 3 mm diameter glassy carbon (GC) disc was used as the working electrode substrate. The catalytic inks were prepared by dispersing 10 mg of catalyst powder in a mixture composed of ultra-pure water (500 µL), isopropanol (500 µL) and Nafion® (100 µL), followed by ultrasonication for 10 min. 3 µL of the dispersed catalyst were deposited onto the surface of the working electrode (catalyst loading of 0.38 mg cm$^{-2}$) and the deposit was dried under $N_2$ flow.

The measurements were conducted in a $N_2$ saturated 1 mol L$^{-1}$ KOH (86.5%, VWR) aqueous electrolyte. Firstly, cyclic voltammograms were recorded in a nitrogen-saturated electrolyte from 0.15 to 1.35 V vs RHE at a scan rate of 50 mV s$^{-1}$. Then the activity of catalysts towards HER was examined by recording linear sweep voltammograms from 0.5 to -0.6 V *vs.* RHE at a scan rate of 5 mV s$^{-1}$ by applying a rotating rate of 1600 rpm to the RDE. All measurements were IR-drop corrected by determining the cell resistance using electrochemical impedance spectroscopy measurements (EIS) between 0.1 Hz-100 KHz. Spectra were acquired in the capacitive region with a Solartron SI 1287 electrochemical interface and an SI 1260 impedance/gain-phase analyser.

3. **Results and discussion**

All the here-investigated MXenes were synthesized from the same initial MAX phase, $Ti_3AlC_2$ powder, as precursor. The synthesis conditions of the different MXenes are summarized in Table 1 and the different separation/washing steps are described in the experimental part and depicted in Scheme 1.

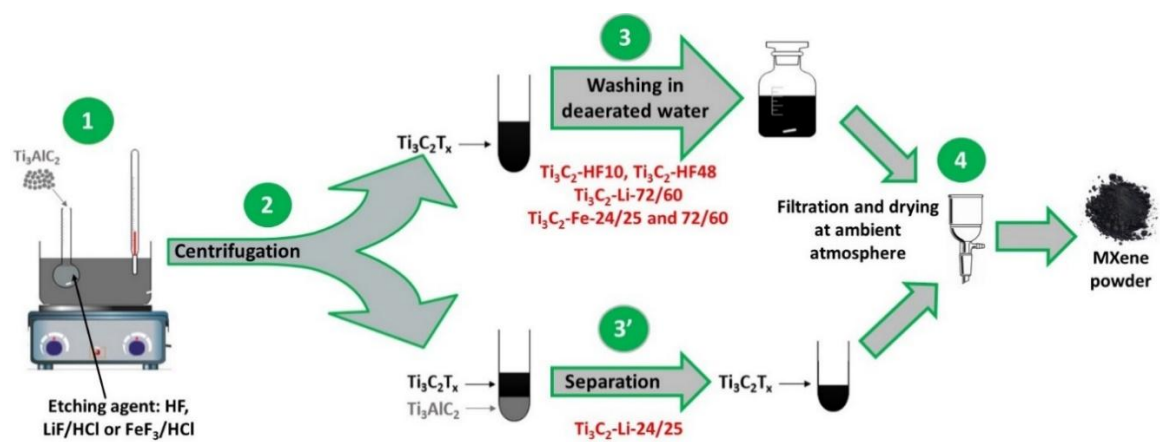

**Scheme 1.** Schematic representation of various MXenes syntheses methods used in this work.

**Table 1.** Summary and labels of synthesis conditions employed to obtain different $Ti_3C_2T_x$ powders. In each case, 0.5 g of $Ti_3AlC_2$ powders and 10 mL of etching solution were used.

| MXene | Labelled condition | Etching agent (10 mL) | Synthesis parameters | Initial F/Al atomic ratio* |
|---|---|---|---|---|
| $Ti_3C_2$-HF10 | soft | HF 10 wt.% | 24 h – 25 °C | 21 |
| $Ti_3C_2$-HF48 | harsh | HF 48 wt.% | 24 h – 25 °C | 98 |
| $Ti_3C_2$-Li-24/25 | soft | 0.8 g of LiF in 9 M HCl | 24 h – 25 °C | 12 |
| Ti$_3$C$_2$-Li-72/60 | harsh | 0.8 g of LiF in 9 M HCl | 72 h – 60 °C | 12 |
| Ti$_3$C$_2$-Fe-24/25 | soft | 1.716 g of FeF$_3$ in 9 M HCl | 24 h – 25 °C | 12 |
| Ti$_3$C$_2$-Fe-72/60 | harsh | 1.716 g of FeF$_3$ in 9 M HCl | 72 h – 60 °C | 12 |

* Atomic ratio between the fluorine contained in the etching agent and the Al in Ti$_3$AlC$_2$.

Three different etching solutions were used. They are:

i) HF with two different concentrations, *viz.* 10 and 48 wt.% (Table 1).

ii) LiF/HCl. This method, first reported in 2014 [8], resulting in the *in-situ* formation of HF, was recently improved by Alhabeb et al [43], with the design of the MILD method (minimally intensive layer delamination) allowing the production of single, as well as large flakes with low amounts of structural defects. Beyond the fact that this method avoids using hazardous HF, the delamination of sheets is favored because of the simultaneous insertion of Li cations and water molecules between sheets during the etching step [47]. In the here reported work, the MILD conditions, *i.e.* exfoliation of Ti$_3$AlC$_2$ at 25 °C during 24 h, was reproduced (sample labelled Ti$_3$C$_2$-Li-24/25). For comparison, a harsher synthesis - 60 °C for 72 h, was investigated (sample labelled Ti$_3$C$_2$-Li-72/60 – Table 1). Increasing both etching temperature and duration was deliberate in order to exacerbate differences between their properties.

iii) FeF$_3$/HCl. As shown in our previous work [36], LiF can be substituted by iron fluoride, FeF$_3$, as the etching agent. These samples - labelled Ti$_3$C$_2$-Fe-24/25 and Ti$_3$C$_2$-Fe-72/60 - were prepared using identical F/Al atomic ratios, temperatures and times (see Table 1) used for the Ti$_3$C$_2$-Li-24/25 and Ti$_3$C$_2$-Li-72/60 samples, respectively.

*3.1 XRD characterization*

The X-ray diffraction (XRD) patterns of the different samples are plotted in Fig. 1. With one exception, all of the etching conditions resulted in the total loss of $Ti_3AlC_2$ peaks – at 2θ of 9.54, 19.18 and 38.92° (see Fig. S1 of ESI) - from the XRD patterns. For the $Ti_3C_2$-Fe-24/25 sample, a very small amount of unreacted MAX phase is observed. The absence of MAX phase peaks in the XRD patterns is taken as direct evidence for the total etching of Al. This result is further confirmed by Inductively Coupled Plasma - Optical Emission Spectrometry (ICP-OES) analysis. The ICP results reported in Table 2 clearly show that the highest number of Al per 3 Ti atoms is only of 0.08. Furthermore, according to the XRD patterns, and with one exception, we see no evidence for crystallized secondary phases, such as TiC, $TiO_2$ or $AlF_3$, that are sometimes observed. In the $Ti_3C_2$-Fe-72/60 sample the presence of TiC peaks was identified.

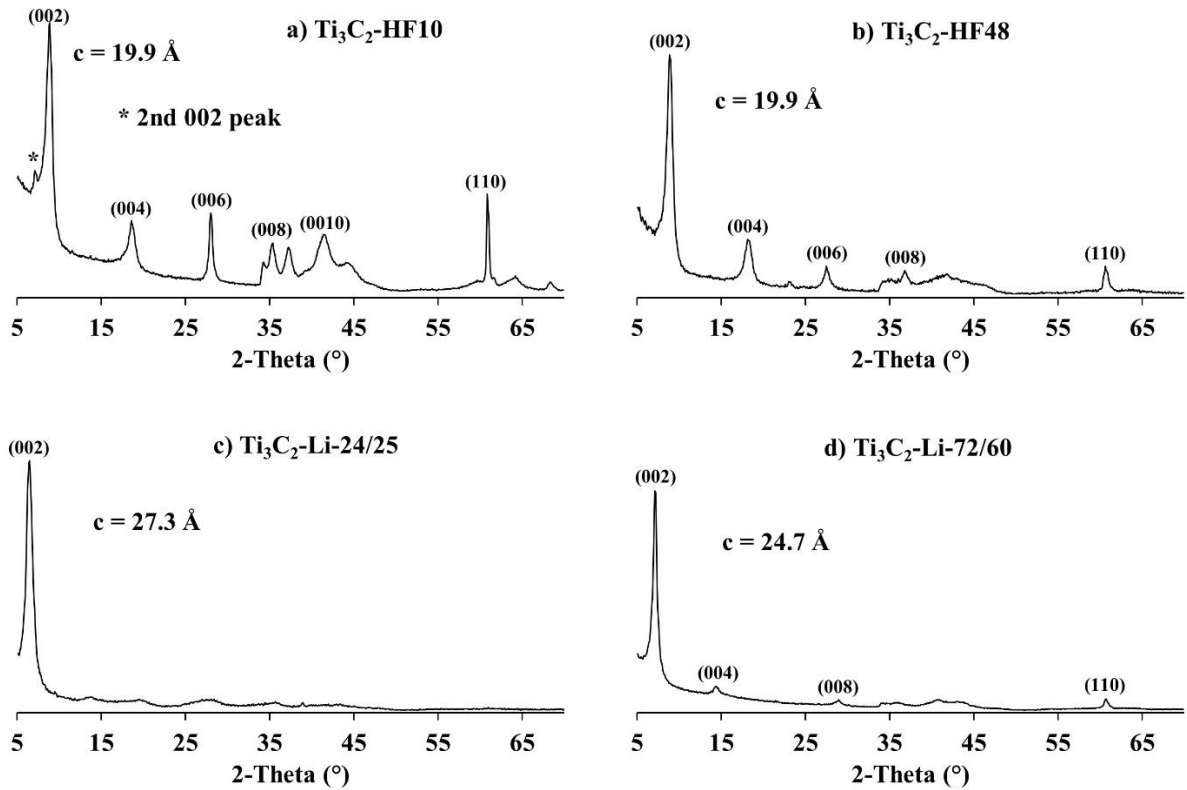

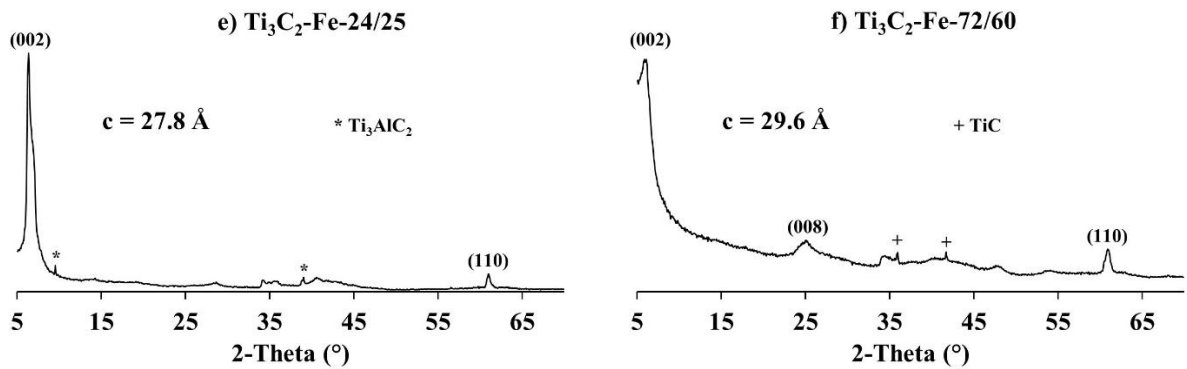

**Fig. 1.** XRD patterns of different $Ti_3C_2T_x$ MXenes. For the sake of clarity, only the (00*l*) and (110) peaks of MXenes are indexed. Note presence of unreacted MAX phase peaks in e) and TiC in f).

**Table 2** ICP analyses of $Ti_3C_2T_x$ final powders prepared with different etching agents. The content of each element is normalized to 3 Ti atoms. The final amount of MXene obtained from 0.5 g of $Ti_3AlC_2$ by the different synthesis methods is reported in last column.

| MXene | ICP-OES | | | | Amount of MXene after synthesis |
|---|---|---|---|---|---|
| | Ti | Al | Li | Fe | (g) |
| $Ti_3C_2$-HF10 | 3 | <0.01[a] | - | - | ≈ 0.40 |
| $Ti_3C_2$-HF48 | 3 | 0.08 | - | - | ≈ 0.51 |
| $Ti_3C_2$-Li-24/25 | 3 | 0.01 | 0.04 | - | ≈ 0.10 |
| $Ti_3C_2$-Li-72/60 | 3 | <0.01[a] | 0.02 | - | ≈ 0.45 |
| $Ti_3C_2$-Fe-24/25 | 3 | 0.04 | - | 0.03 | ≈ 0.40 |
| $Ti_3C_2$-Fe-72/60 | 3 | 0.05 | - | 0.06 | ≈ 0.22 |

[a] Below the detection limit

The XRD pattern of the $Ti_3C_2$-HF48 powders is similar to those previously reported using the same preparation conditions [33]. It is well-established that the substitution of Al by terminal groups during the etching step involves a shift of (00*l*) peaks of the initial MAX phase towards lower diffraction angles leading to higher *c* lattice parameter value ($c$ = 18.5 Å for $Ti_3AlC_2$). The *c* parameter value of 19.9 Å obtained for the $Ti_3C_2$-HF48 material is the lowest value obtained herein, indicating that *no* water is intercalated between sheets for this material. For the $Ti_3C_2$-HF10 powders, two (002) peaks can be clearly identified: a more intense one corresponding to a *c* parameter value of 19.9 Å and a less intense one corresponding to a *c* parameter value of 25.5

Å. The latter indicates the presence of some water between the sheets and is characteristic of the interstratification phenomenon observed in clay materials, as described in our previous work [46]. This hydration heterogeneity involves different layer-to-layer distances in the multilayer particle. The observation of two distinct peaks is also characteristic of partial segregation, *i.e.*, wherein a succession of layers having the same hydration state is favored over a random one [46].

As observed in Fig. S2, before the washing step (step 3 of Scheme 1), the $Ti_3C_2$-HF10 sample is mainly composed of stacks having an interlayer spacing leading to a *c* parameter value of 25.5 Å (Fig. S2a) revealing higher amounts of water layers intercalated in the MXene structure. It is well-known that the presence of cations is required to stabilize water molecules between the sheets since the MXene surface is negatively charged [35,46]. So, in the present case, it is very likely that $H_3O^+$ (coming from the acid of the etching agent) stabilizes the interlayer water molecules. The stabilization of water molecules with $H_3O^+$ was already reported [47]. The washing step, performed at low concentration - 0.5 g in 350 mL of deaerated water - has a crucial role for this sample since it removes these cations, probably by osmotic pressure, leading to a partial dehydration of the MXene. Consequently, washing results in a decrease in the *c* parameter value and the aforementioned segregation phenomenon is observed. In contrast, the XRD patterns of the $Ti_3C_2$-HF48 powders (Fig. S2c and d) evidence a negligible effect of the washing step on c. The *c* parameter at 19.9 Å does not evolve with washing, indirectly demonstrating the absence of water and cations between sheets. Thus, in contrast to the $Ti_3C_2$-HF48 sample, the $Ti_3C_2$-HF10 multilayers exhibit a real ability for the insertion of few water molecules between their sheets. Nevertheless, once the water layers are removed, it is not possible to increase again the d-spacing since no cations are present between the layers allowing their stabilization.

In the $Ti_3C_2$-Li-24/25 case, a separation step by centrifugation (step 3' – Scheme 1) was required to obtain a MXene free from the initial MAX phase. A high fraction of unreacted MAX phase always remained in the sediment (see Fig. S3) and the complete separation of the MAX and MXene was difficult. It is for this reason that the fraction of obtained MXene for this etching protocol (≈ 0.1 g) is considerably smaller than the other methods (see Table 2). Nevertheless, as proposed by Alhabeb et al [43], the sediment can be reused to undergo the synthesis protocol. In comparison with the (002) peaks, the intensities of other (00*l*) peaks are so low that a higher magnification of the diffractogram is required to reveal them (Fig. S3b). The asymmetry of these (00*l*) peaks combined with the position of the (002) peak corresponding to a c parameter value of 27.3 Å, value between 25 Å (one water layer) and 31 Å (two water layers), reveals a random distribution of interlayer spacing having one water layer or two water layers stabilized by the $Li^+$ cations [35,46], characteristic of interstratified MXenes [46].

The XRD pattern of $Ti_3C_2$-Li-72/60 sample (Fig. 1) is close to that of $Ti_3C_2$-Li-24/25 sample with a *c* parameter value of 24.7 Å. As expected [8,36], both MXenes have intercalated water molecules between the sheets contrary to the $Ti_3C_2$-HF48 (19.9 Å) powders for which no water was present between the layers. It should be noted that a LiF impurity is sometimes observed in the XRD patterns (Fig. S3b). Nevertheless, the LiF amount is quite low as confirmed by the Li content determined from ICP analyses (Table 2). Finally, the use of harsher conditions ($Ti_3C_2$-Li-72/60 material) makes it possible to obtain a higher amount of MXene (≈ 0.45 g – Table 2) in a one-pot synthesis compared to the $Ti_3C_2$-Li-24/25 sample and does not require an additional separation step (like in step 3' of Scheme 1).

For the $FeF_3$-etched samples, the characteristic MXene diffraction are also observed and traces of the $Ti_3AlC_2$ phase are present for the $Ti_3C_2$-Fe-24/25 sample (Fig. 1e and f).

Interestingly, whereas the etching conditions are similar for $Ti_3C_2$-Fe-24/25 and $Ti_3C_2$-Li-24/25 samples, the etching of Al is clearly favored when $FeF_3$/HCl is used since the fraction of unreacted MAX phase is quite low and the separation step (step 3' of Scheme 1) is not required. Furthermore, the obtained MXene amounts in a one-pot synthesis is four times higher: 0.4 g and 0.1 g for $Ti_3C_2$-Fe-24/25 and $Ti_3C_2$-Li-24/25 samples, respectively (Table 2). Consequently, Al etching is not only due to the *in-situ* generated HF (the initial fraction of F is similar for both samples) but also the cations present. In this case, Fe also plays a major role in the etching process and this will be discussed in section 3.4 of this manuscript. The asymmetric shape of the XRD peaks obtained when Fe is used, also indicates an interstratification phenomenon. The *c* parameter values of 27.8 Å and 29.6 Å calculated for $Ti_3C_2$-Fe-24/25 and $Ti_3C_2$-Fe-72/60 MXenes, respectively, are higher than those obtained when starting with HF or LiF in the etching solutions, suggesting the simultaneous insertion of water molecules and Fe cations between the MXene sheets. The amount of Fe present, however is quite low (Table 2) but previous works showed that a low amount of cations is enough to stabilize water layers (for example: 0.08 Mg for 3 Ti) [47].

For the $Ti_3C_2$-Fe-72/60 sample, only 0.22 g of MXene was obtained, indicating that a significant amount of the MAX phase and/or the synthesized MXene dissolves during the process. At 0.45 g, the remaining MXene amount for the $Ti_3C_2$-Li-72/60 sample is more than double. Thus the $FeF_3$-etching agent is significantly harsher and results in over-etching. The low amount of TiC observed as a secondary phase in the XRD pattern recorded for $Ti_3C_2$-Fe-72/60 sample (Fig. 1) confirms the over-etching. Indeed, TiC is a small impurity in the initial MAX phase which is not affected during the etching process and thus remains after the synthesis. Except for $Ti_3C_2$-Fe-72/60 sample, the fraction of TiC in the global sample is very low and is not detected by XRD. In contrast, for the latter, since a relative large fraction of MAX/MXene dissolves, the fraction of TiC in the global sample increases explaining why it is observed by XRD.

*3.2 Raman and XPS spectroscopies*

The XPS spectra of Ti 2p, O 1s, C 1s and F 1s of the different samples are plotted in Fig. 2 and the fitting results are reported table S1 to S7. XPS is currently an essential tool to understand the surface chemistry of MXenes and several recent papers have reported on the spectral decomposition of the different photopeaks associated with MXenes [48–50]. As noted above, one of the aims of the present work is to provide a better methodology to decompose experimental XPS signals, thanks to the synthesis of $Ti_3C_2T_x$ samples with different surface compositions, and to propose a better quantification of the terminal groups.

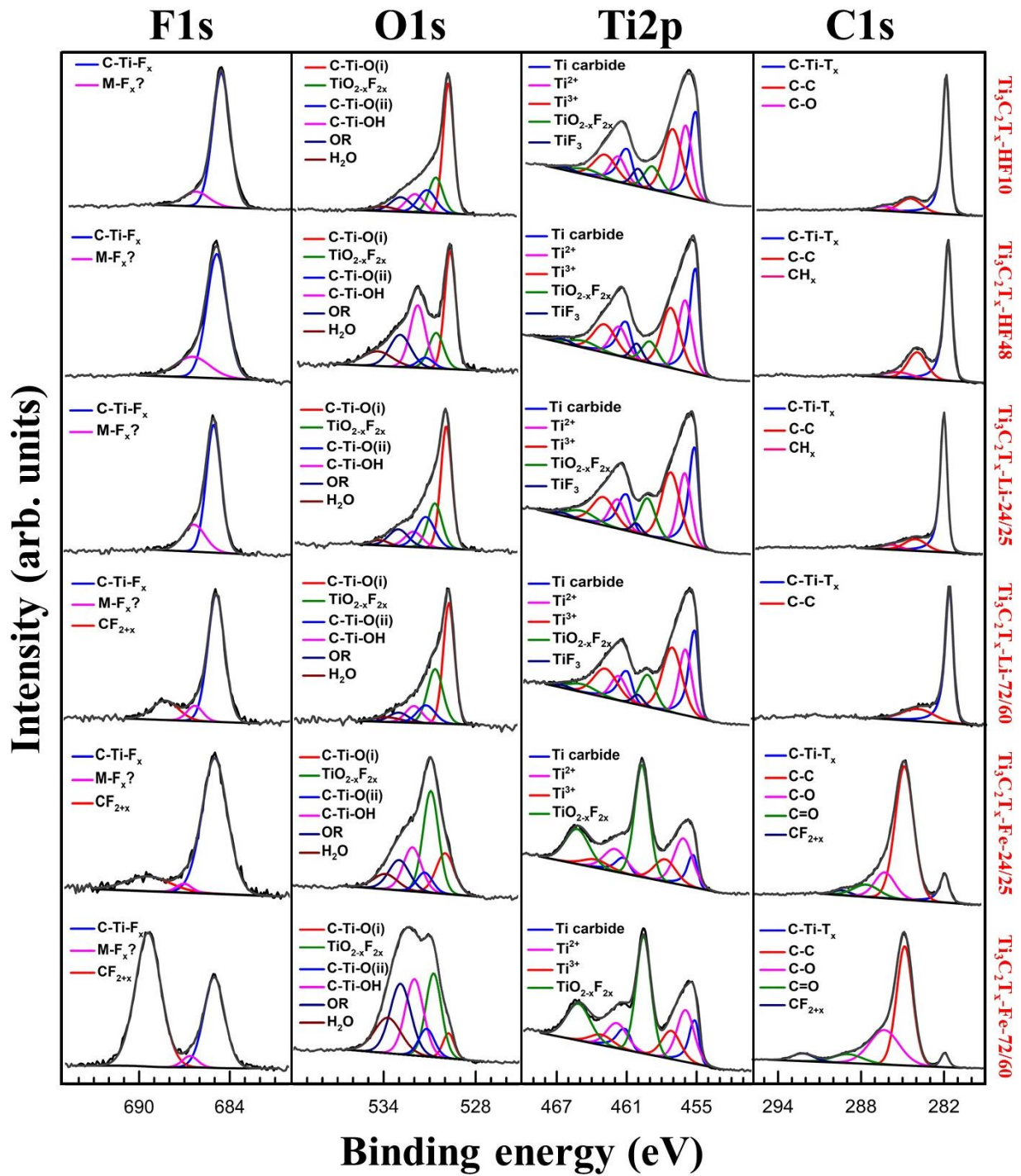

**Fig. 2.** XPS high resolution spectra of F 1s, O 1s, Ti 2p and C 1s regions for the different $Ti_3C_2T_x$ samples whose labels are listed on right hand side.

In the Ti 2p region, it is now well-established that the 3 doublets of asymmetric peaks centered at *ca.* 455.0 eV (only the first value of the doublet is reported here), 455.8 eV and 457.0 eV and labeled as Ti carbide, $Ti^{2+}$ and $Ti^{3+}$, respectively, are related to the MXene structure: C-Ti-C in the core of the layer, C-Ti-O, C-Ti-F and C-Ti-OH corresponding to Ti atoms bonded to both C atoms and terminal groups. Nevertheless, the assignation of each peak to a specific bond is not clear since numerous discrepancies can be found in the literature [48–50]. Moreover, due to their closeness in terms of binding energy, the decomposition of the 3 peaks is not obvious since different combinations of peak positions and full width at half maxima (FWHM) can provide a good refinement while drastically changing the area ratio between peaks. Nevertheless, for quantification purposes, we can undoubtedly attribute these 3 contributions to MXene Ti atoms.

In the pioneering XPS work done by Halim *et al.* [48], the Ti atoms bonded to F-terminal groups was shown to be responsible for the appearance of the band centered at *ca.* 460.0 eV. By correlating the area of this latter peak with the one of C-Ti-F in the F 1s spectra, the obtained F/Ti ratio is inconsistent (see Tables S2 to S7 and the F/Ti ratio determination in part III of ESI). For example, the F/Ti ratios of ≈ 11 and 7 are found for the $Ti_3C_2$-HF48 and $Ti_3C_2$-HF10 samples, respectively. This is impossible since in the $Ti_3C_2T_x$ formula group, x is considered to be 2 or less [48]. Moreover, the magnitude of this peak is similar for $Ti_3C_2$-HF48 and $Ti_3C_2$-HF10 (table S2 to S3) whereas it is clear that $Ti_3C_2$-HF48 have more F terminal groups as discussed below. Thus, it is clear that F terminal groups contributions should be involved in the three above mentioned contributions assigned to the MXene Ti atoms as proposed by Schultz *et al.* [50] and Persson *et al.* [49]. The peak centered at *ca.* 460.0 eV and labelled in the present work as $TiF_3$ is therefore herein attributed to the formation of a minor secondary phase, $TiF_{3\pm x}$, formed during the etching process. To confirm this conclusion, we measured the Ti 2p region of a commercial $TiF_3$ powder. A peak was obtained at a B.E. of 460.5 eV (Fig. S4a) which is close enough to our 460.0 eV peak.

Furthermore, the doublet at *ca.* 459.0 eV (Fig. 2 and table S2 to S7) in the Ti 2p region was assigned to the presence of $TiO_{2-x}F_{2x}$. In general, the peak corresponding to $TiO_2$ is centered at *ca.* 458.8 eV [51]. However, the introduction of small amounts of F in this structure, resulting in $TiO_{2-x}F_{2x}$, leads to a shift of this peak towards higher B.E. values (*ca.* 459.1 eV) [51]. Thus, the oxidation observed for the different samples can be assigned to the formation of a $TiO_{2-x}F_{2x}$ oxyfluoride. This comment notwithstanding, one cannot exclude that this shift toward a higher B.E. simply results from a stronger electronic interaction of $TiO_2$ with the MXene conductive substrate.

The magnitude of this contribution is particularly high for the Fe samples showing that these materials are the most oxidized ones (Fig. 2). Therefore, this photopeak can be safely assigned to an oxidation of the MXene surface resulting in the formation of a fluorinated $TiO_2$ viz. $TiO_{2-x}F_{2x}$. Raman spectroscopy, discussed in the next section, confirms this conclusion. The fraction of oxidized Ti atoms was determined by calculating the ratio between the area of the peak corresponding to $TiO_{2-x}F_{2x}$ (459.1 eV) and all peaks involving chemical bonds formed with the Ti atoms (Table 3).

Peaks in the O 1s region were assigned based on reported studies [52]. The peaks centered at *ca.* 529.7-529.8 eV and at *ca.* 531.2-531.3 eV labelled as C-Ti-O (i) and C-Ti-O (ii) in our work can be attributed to –O terminal groups bonded to Ti atoms at two different crystallographic sites, previously described as bridging and A sites, respectively [50]. The peak at *ca.* 531.8-532.0 eV can be ascribed to hydroxyl terminal group bonded to Ti atoms. Lastly, the peak centred at *ca.* 533.7-534.1 eV was attributed to water molecules (inserted between the atomic sheets or at the surface of the multi-layered particles). The peak centered at *ca.* 530.6-530.8 eV is attributed to $TiO_{2-x}F_{2x}$. Importantly, the intensity of this peak increases as that of the peak attributed to $TiO_{2-x}F_{2x}$ in the Ti 2p region also increases, particularly for the oxidized Fe samples, thus confirming our assignment. Finally, the contribution at *ca.* 533 eV corresponds to O bonded to adventitious carbon (labelled OR in Fig. 2) as proposed by Schultz *et al.* [50]. Nevertheless, we cannot exclude that a small fraction of this adventitious O is also present in the peak assigned to hydroxyl groups at *ca.* 531.8-532.0 eV.

In the C 1s region, the peak at *ca.* 281.9 eV can be undoubtedly assigned to $C-Ti-T_x$ moieties [48–50]. The others contributions (Table S2 to S7) are ascribed to the contamination of the MXene surface by the atmosphere [48,50]. Nevertheless, the high magnitude of the C-C contribution in the most oxidized samples ($Ti_3C_2$-Fe-24/25 and $Ti_3C_2$-Fe-72/60) can be also partly due to the formation of amorphous C during the etching process due to oxidation and over-etching (Fig. 2). The formation of amorphous carbon during MXene oxidation process was already reported in the literature [53].

In the F 1s region, the major contribution at *ca.* 685.0 eV can be assigned to C-Ti-F groups. Nevertheless, F atoms belonging to $TiO_{2-x}F_{2x}$ and $TiF_3$ compounds also contribute to this peak. Indeed, the F 1s peak for $TiO_{2-x}F_{2x}$ is 684.6 eV [51]. We measured a B. E. of 685.1 eV for $TiF_3$ (Fig. S4a). The low intensity peak at 686.5 eV is probably related to an unknown fluorinated secondary phase or to F atoms directly bonded to MXene C atoms exposed due to the dissolution of some of Ti atoms. The others peaks at higher B.E can be assigned to the formation of F polymeric chains involving $CF_2$ or $CF_3$ groups [54]. As observed, this photopeak is present almost exclusively on MXenes prepared with the $FeF_3$/HCl etching solution which are the most oxidized samples. This is not surprising since the oxidation process involves the breaking of Ti-C bonds allowing for the formation of fluorinated polymeric C chain thanks to the presence of F atoms in the etching environment.

**Table 3** Results of XPS quantification for all MXenes.

| Sample | MXene Composition | Sum of terminal groups | Fraction of oxidized Ti atoms |
|---|---|---|---|
| $Ti_3C_2$-HF10 | $Ti_3C_{2.08}O(i)_{0.81}O(ii)_{0.23}(OH)_{0.19}F_{0.61}$ | 1.85 | 9.8% |
| $Ti_3C_2$-HF48 | $Ti_3C_{2.03}O(i)_{0.43}O(ii)_{0.06}(OH)_{0.30}F_{1.34}$ | 2.13 | 9.4% |
| $Ti_3C_2$-Li-24/25 | $Ti_3C_{1.94}O(i)_{0.71}O(ii)_{0.23}(OH)_{0.17}F_{0.75}$ | 1.86 | 14.3% |
| $Ti_3C_2$-Li-72/60 | $Ti_3C_{1.86}O(i)_{0.81}O(ii)_{0.19}(OH)_{0.16}F_{0.70}$ | 1.86 | 13.3% |
| $Ti_3C_2$-Fe-24/25 | $Ti_3C_{2.03}T_x$ | nd[a] | 48.1% |
| $Ti_3C_2$-Fe-72/60 | $Ti_3C_{1.99}T_x$ | nd[a] | 46.7% |

[a] nd = not determined

Based on our aforementioned assignments of the various XPS peaks, a quantification of the MXene surface composition was attempted (Table 3) using the results of the fits reported in Tables S2 to S7. To this purpose, several assumptions were made: (i) F attributed to the unknown fluorinated phase was not considered as an integral part of the MXene, (ii) amount of F atoms bonded to Ti atoms in the $TiF_3$ impurity was subtracted from the C-Ti-$F_x$ contribution at 685 eV considering 3 F atoms for 1 Ti, the amount of Ti in $TiF_3$ being deduced from the $TiF_3$ contribution centered at *ca.* 460.0 eV in the Ti 2p region (Fig. 2), (iii) F atoms belonging to the $TiO_{2-x}F_{2x}$ moiety in the C-Ti-F contribution at 685 eV as well as adventitious O in the O 1s region and involved in the C-Ti-(OH)$_x$ groups were neglected, (iv) the low amounts of chlorine, Cl, atoms (Table S1) were ignored, even if some are terminal groups.

Based on these different assumptions and taking into account the fittings uncertainties (close to 10%), the obtained results should be taken with a grain of salt. Nevertheless, the obtained compositions of the different MXenes are logical (Table 3), validating our approach. Indeed, in every case, a Ti/C ratio of 3/2 is obtained. As importantly, the global number of moles of T groups is close to 2 as expected [48]. It is worth noting here, quantification of the terminal groups for the FeF$_3$-etched samples was too complicated and is thus not reported in Table 3. This was mainly due to the high degree of oxidation of these surfaces. Indeed, the fitting of the O 1s region was strongly impacted by the formation of oxygenated carbonaceous species caused by the corrosive etching. The high fraction of Ti oxyfluoride complicates the quantification of the F terminal groups since the x value in $TiO_{2-x}F_{2x}$ could not be unambiguously determined. Nevertheless, the consistent Ti/C ratio, obtained from spectral decomposition of the Ti and C peaks indirectly confirms the validity of our peak assignments and the formation of MXenes with this etching agent.

According to Table 3, the compositions obtained for the Ti$_3$C$_2$-HF10, Ti$_3$C$_2$-Li-24/25 and Ti$_3$C$_2$-Li-72/60 samples were quite close. The F-content of the Ti$_3$C$_2$-HF48 composition, on the other hand, was roughly double that of the other samples (Table 3 and Table S1). The higher amount of F atoms in the Ti$_3$C$_2$-HF48 sample is not surprising given the higher F content in the etching solution (Table 1) [6]. Somewhat surprisingly, the amount of F atoms in all Li-samples is close to that of the Ti$_3$C$_2$-HF10 samples despite the fact that almost twice as much F was used to synthesize the latter (Table 1).

The fraction of hydroxyl groups among the oxygenated species is higher in the Ti$_3$C$_2$-HF48 sample in comparison with Ti$_3$C$_2$-HF10, Ti$_3$C$_2$-Li-24/25 and Ti$_3$C$_2$-Li-72/60 samples. As discussed below, this conclusion is confirmed by Raman spectroscopy. As already shown, the fraction of O/OH on the MXene surface depends on pH [55]. At low pH the favoured termination is OH; increasing the pH leads to dihydroxylation of the OH into O. Our synthesis was carried out under acidic conditions, favouring hydroxyl groups but the different steps of centrifugation with water (see experimental part) increase the pH until 5 leading to a partial dihydroxylation and the formation of O terminal groups on the MXene surfaces as proposed by Natu et al [55]. Nevertheless, the dihydroxylation for $Ti_3C_2$-HF48 sample appears to be slower. As observed Fig. S2, after the centrifugation steps and before the washing step, water layer can be inserted between the layers for the $Ti_3C_2$-HF10 sample which is not the case for the $Ti_3C_2$-HF48 sample. Thus, for the latter, the dihydroxylation, involving water molecules, is disfavoured [55]. Even if it is to a lesser extend for the $Ti_3C_2$-HF48 sample, this finding (table 3) confirms that O is favoured as terminal group in comparison with OH in our synthesis conditions and in agreement with studies reported by Persson et al. [49] and Hope et al. [56].

Moreover, as previously mentioned, the F terminal groups are bonded preferentially on the A-sites [49,50]. Thus, it is not surprising that, when the amount of F terminal groups increases as in the case of $Ti_3C_2$-HF48, a smaller fraction of O atoms are on the A-site (C-Ti-O (ii)). This result confirms our assignation and the theoretical work of the literature [49,50].

Finally, the fraction of oxidized Ti atoms (Table 3) indicates that the MXenes prepared with the $FeF_3$/HCl etchant are highly surface oxidized and no significant difference was observed between both the two samples. This result further confirms that this etching agent is more corrosive than LiF/HCl etching agent as discussed in the XRD characterization section.

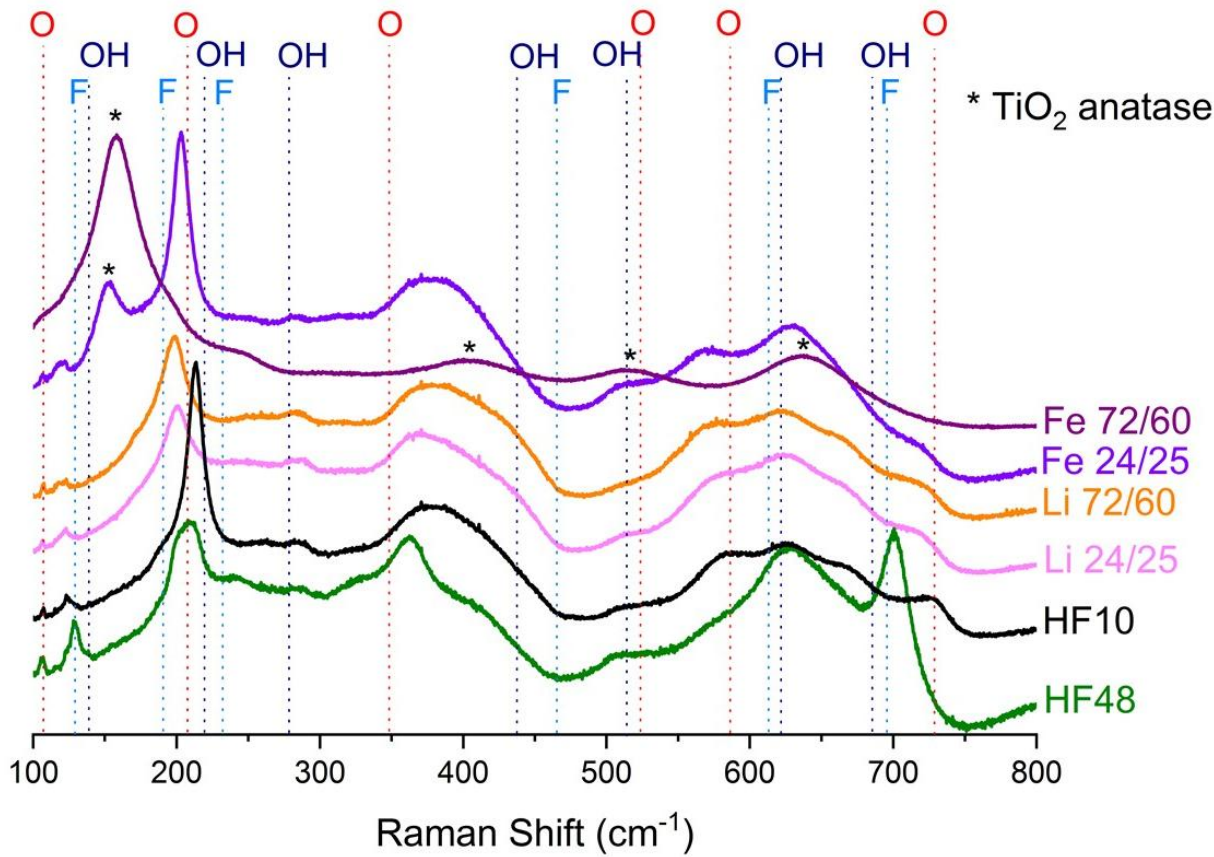

**Fig. 3.** Raman spectra of the as-synthesized MXenes. The spectra are shifted vertically for clarity.

Raman spectra obtained for all samples are plotted in Fig. 3. According to the computational study of Hu *et al.* [57], there are several Raman active vibrational modes in the observed range for the three $Ti_3C_2T_x$ with homogeneous terminal groups, namely $Ti_3C_2O_2$, $Ti_3C_2F_2$ and $Ti_3C_2(OH)_2$, and the chemical nature of the terminal groups greatly affects the vibrational frequencies of the Raman modes. In Fig. 3, the positions of the calculated frequencies are indicated by doted lines and marked with an O, F or OH label, respectively for $Ti_3C_2O_2$, $Ti_3C_2F_2$ and $Ti_3C_2(OH)_2$. It can be seen that these calculated frequencies roughly match with the observed ones and that, except for the $Ti_3C_2$-Fe-72/60 sample, the different spectra are indicative of the formation of $Ti_3C_2T_x$ MXene. Indeed, deviations observed from the calculated values may be explained since calculations are based on $Ti_3C_2T_2$ MXene with homogeneous terminations and only considering a monosheet [57].

The intensities of the two bands at 128 and 700 cm$^{-1}$, assigned to $Ti_3C_2F_2$, were found to be relatively higher for the $Ti_3C_2$-HF48 sample than the others, confirming the higher F content in good agreement with the XPS analyses (Table 3 and S1).

Based on the theoretical results of Hu *et al.* [57], the frequency associated with the out-of-plane stretching vibration of surface Ti in bare $Ti_3C_2$ at 228 cm$^{-1}$ is shifted to 218 and 208 cm$^{-1}$, respectively with -OH or -O as terminal groups and to 190 cm$^{-1}$ with -F as terminal groups. The position of this band in the spectra will be therefore greatly dependent on the relative abundance of the different terminal groups. In Fig. 3, the corresponding band is centered at about 212 cm$^{-1}$ for the $Ti_3C_2$-HF10 sample. For the F-rich sample ($Ti_3C_2$-HF48), a shoulder ≈ centered at 195 cm$^{-1}$ is clearly observed again consistent with the higher F content of this sample. Nevertheless, this band is also downshifted for samples prepared with the LiF/HCl etching agent, whereas the amount of F is close to that of $Ti_3C_2$-HF10 (Table 3). Lioi *et al.* [58] also observed the downshift of this band when comparing MXenes obtained using HF and LiF/HCl etching agents. They state that small changes in the relative abundance of the various $T_x$ groups may have a strong effect in individual MXene sheets and therefore significantly affect the Ti-C bond lengths and consequently Raman modes. Inter-layer species (water, cations), present in the LiF etched samples, should also affect the electron density drawn from the Ti-C bonds resulting in harder/softer Raman modes. This Raman band (*ca.* 200 cm$^{-1}$) was also shown to be affected by the restacking properties of the materials [59]. Besides, we cannot exclude that Cl atoms observed in small amounts in LiF and FeF$_3$-etched samples (Table S1) may also affect this out-of-plane stretching vibration mode in the same way as F atoms do.

The band intensity associated with the -O terminal groups (730 cm$^{-1}$) increase as the HF concentration decreases indicating that more -O terminal groups are present. The intensity of bands associated with the –OH terminal groups (622 and 285 cm$^{-1}$) does not evolve significantly, however. Thus, the O/OH ratio of the Ti$_3$C$_2$-HF48 sample is lower than that of all other samples. This result is also in line with the XPS study and validates both our XPS and Raman assignments.

The Raman spectra of both samples etched with the LiF/HCl solution are similar to the ones etched with 10% HF (Ti$_3$C$_2$-HF10) except for the already discussed shift of the band associated to the out-of-plane stretching vibration of surface Ti atoms (*ca.* 200 cm$^{-1}$). This is also in agreement with the XPS data showing no significant differences between the surface compositions of these three samples (Table 3).

Though the XRD patterns of the Fe-samples do not show the presence of crystallized TiO$_2$, Raman and XPS spectroscopies clearly indicate the partial oxidation of these samples. Indeed, the formation of anatase TiO$_2$ (or TiO$_{2-x}$F$_{2x}$) can be observed in the Raman spectra (Fig. 3), particularly for Ti$_3$C$_2$-Fe-72/60 sample. The E$_g$ vibrational mode of pure anatase is at 143 cm$^{-1}$ [51]. Here, that mode is shifted to 158 cm$^{-1}$, a shift that could be due to the presence of F in the anatase structure [51]. This conclusion is consistent with our XPS results. The fraction of TiO$_{2-x}$F$_{2x}$ increases under harsher etching conditions (Ti$_3$C$_2$-Fe-72/60).

Referring to Fig. 3, the presence of the MXene spectral signature is clear for the Ti$_3$C$_2$-Fe-24/25 material. For the visible Ti$_3$C$_2$-Fe-72/60 sample the intense TiO$_{2-x}$F$_{2x}$ bands hide the MXene spectral footprint. It is also obvious that this oxidation process occurs primarily on the surface of the MXene as evidenced previously from XPS since the high fraction of $TiO_{2-x}F_{2x}$ at the surface is similar for both samples (Fig. 2 and Table 3) whereas the amount of $TiO_{2-x}F_{2x}$ observed by Raman, with a greater depth of analysis, is clearly different (Fig. 3).

Finally, a hump is also observed between 350 and 450 cm$^{-1}$ for all $Ti_3C_2T_x$ samples. As reported by Hu *et al* [57], this can be explained by the heterogeneity of terminal groups. Indeed, using a simplified $Ti_3C_2O(OH)$ monosheet model, two Raman active modes appear at 364 and 387 cm$^{-1}$ which match well with our results. We thus conclude, not unsurprisingly and in total agreement with previous work, that our MXenes have heterogeneous terminations, as observed from XPS results. Furthermore, the intensity of these bands seems to be strongly affected by the nature of terminal groups as the intensity of the band at *ca*. 364 cm$^{-1}$ considerably increases with the F-content.

*3.3 Micro/macrostructure characterization*

The macrostructure of the MXene is also affected by the chemical composition of the etching agent. As shown by SEM analyses (Fig. S5), using a high HF concentration ($Ti_3C_2$-HF48) leads to multi-layers MXenes separated by large gaps (accordion-like structure) whereas at lower HF concentrations ($Ti_3C_2$-HF10) tightly stacked sheets are formed [43]. The exfoliation process performed using HF at high concentrations involves the formation of more macroscopic defects and thus well-separated multi-layers MXenes [43]. This is due to the formation of $H_2$ gas from the reaction between Al and HF, this reaction being faster when the HF concentration increases thus contributing to expanding the structure. SEM pictures of Li-samples (Fig. S5) show the typical microstructure of a delaminated MXene after restacking during the filtration step [8]. Regardless of the synthesis conditions of the LiF-etched samples, these results point out that the MILD method leads to easier delaminations. Indeed, the delamination can be performed by simply shaking the suspension.

The general microstructure of FeF$_3$-etched samples observed by SEM is similar to that of HF-MXene (Fig. S4). Contrary to LiF-etched samples, here the delamination is not observed by simple manual shaking though the *c* parameter value is higher. Even if some flakes can be isolated by centrifugation as indicated in our previous work [36], this method does not allow for a facile delamination and formation of clays as in the case of LiF/HCl, probably due to surface oxidation. Previous works showed already that the spontaneous delamination in water is only possible with Li cations [47].

From Raman spectroscopy, we concluded that the Ti$_3$C$_2$-Fe-72/60 material is highly oxidized. Using high magnification SEM images (Fig. 4) of that sample confirmed the presence of a large number of surface nanoparticles that are presumably TiO$_2$ anatase. Such nanoparticles are not observed on the non-oxidized samples such as Ti$_3$C$_2$-HF10 (Fig. 4). The presence of these nanoparticles results in a significant increase in the specific surface area values determined by BET method - 75 m$^2$ g$^{-1}$ for Ti$_3$C$_2$-Fe-72/60, and between 2 and 15 m$^2$ g$^{-1}$ for the other samples (Table S8).

EDS analyses were also undertaken on the different samples (Table S8). Taking into account the uncertainties concerning the quantification of light elements, the reported results only allows for a semi-quantitative analysis. The removal of Al in all samples deduced from ICP measurements (Table 2) is confirmed by EDS analyses. As observed by XPS (Table 3 and S1), a higher content of F is observed for Ti$_3$C$_2$-HF48 material compared to the other samples. In agreement with the XPS and Raman analyses, no major differences in compositions are observed between the Ti$_3$C$_2$-HF10, Ti$_3$C$_2$-Li-24/25 and Ti$_3$C$_2$-Li-72/60 samples. The presence of Cl atoms in the MXene prepared with HCl in the etching solution is also confirmed. The O fraction in the FeF$_3$-etched samples is significantly higher than for other samples due to the MXene oxidation, increasing accordingly to the fraction of TiO$_{2-x}$F$_{2x}$ observed by Raman spectroscopy (Fig. 3). The C fraction also increases, particularly for the Ti$_3$C$_2$-Fe-72/60 sample, confirming the formation of an amorphous C phase during etching as demonstrated by XPS analysis. Titanium atoms are indeed partly oxidized and dissolution takes place during the synthesis as stated previously, automatically increasing the fraction of C in the sample.

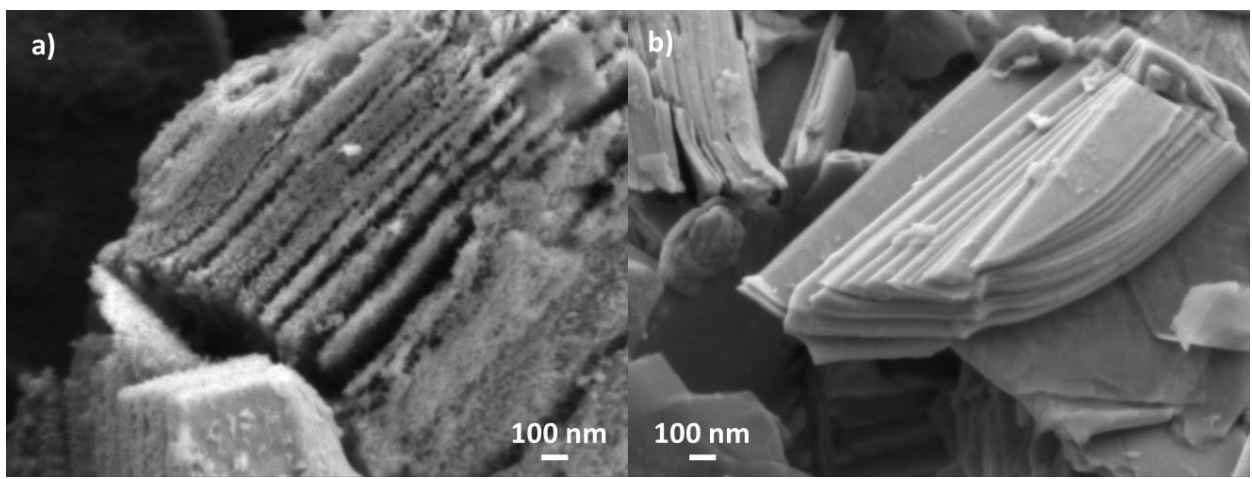

**Fig. 4.** SEM micrograph of, a) Ti$_3$C$_2$-Fe-72/60 sample at high magnification showing the formation of small nanoparticles on the surface of the MXene sheets, b) Ti$_3$C$_2$-HF10 sample, at same magnification, showing the absence of nanoparticles.

Comparing the behavior of the MXenes synthesized using the three different etching agents, only those etched with LiF/HCl allows for a spontaneous delamination by simply manually shaking an aqueous MXene suspension [47]. That is not say that the two LiF-etched samples were different. To highlight a difference, the powders were dispersed in water to form a colloidal suspension and drop-cast onto a SiO$_2$/Si substrate for SEM observation (Fig. 5). From the SEM images, major differences were observed, particularly concerning the lateral size of isolated flakes (or multi-layers flakes with few sheets). Indeed, the harsher etching conditions led to smaller flakes (with lateral sizes < 2 μm) whereas the $Ti_3C_2$-Li-24/25 sample is composed of larger flakes (most with a lateral size between 2 and 10 μm).

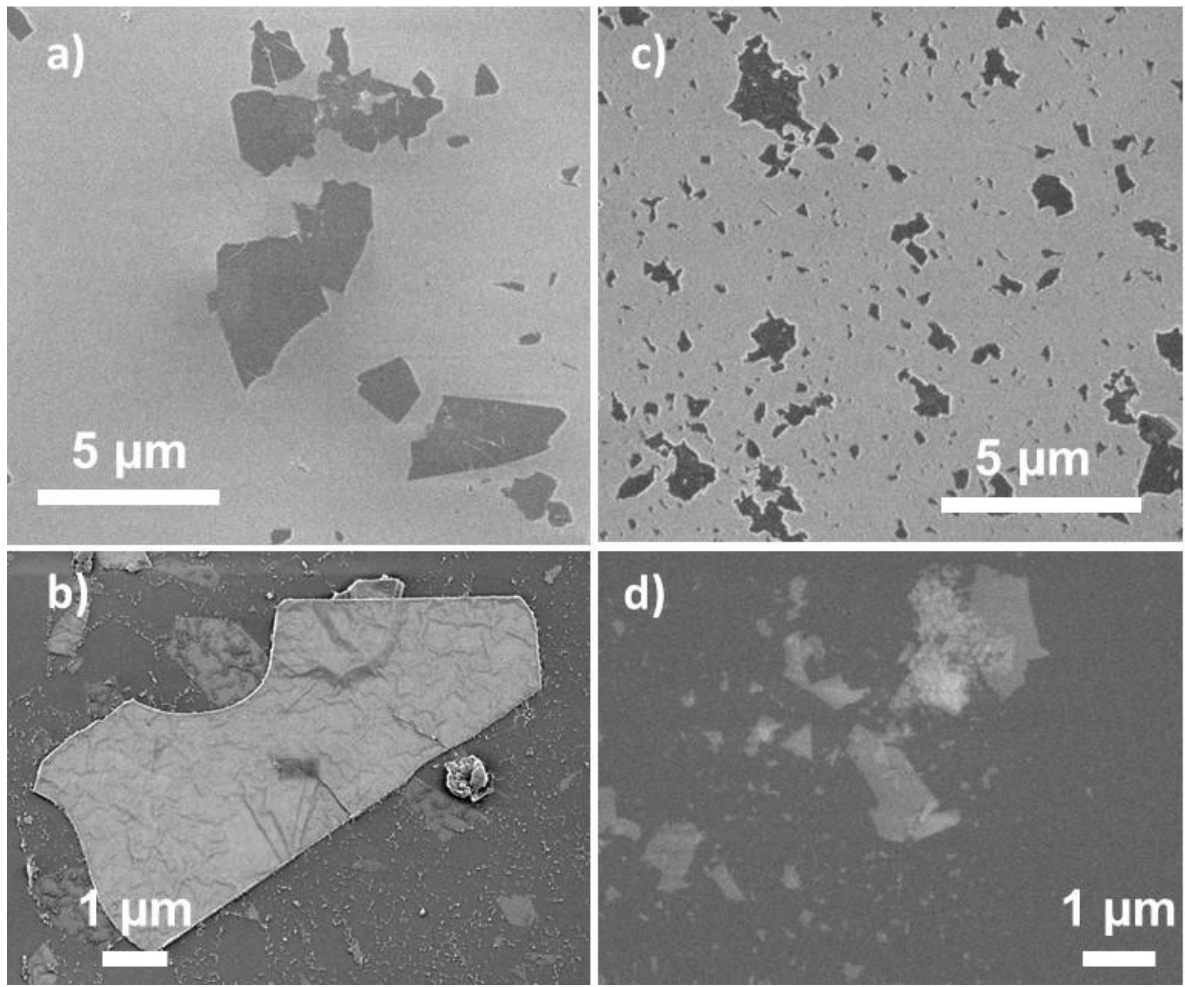

**Fig. 5.** Typical SEM micrographs of flakes collected from colloidal suspension of, a) and b) $Ti_3C_2$-Li-24/25 and c) and d) $Ti_3C_2$-Li-72/60. Top: secondary electron imaging, bottom: backscattered electron imaging.

In addition to reducing the flake size, the harsh conditions (*i.e.* those corresponding to the synthesis of $Ti_3C_2$-Li-72/60) led to more disorganized stacks and defective sheets as evidenced in Fig. 6, where a TEM analysis of representative flakes is shown. Fig. 6a shows different TEM micrographs and corresponding energy-filtered electron diffraction patterns (EFEDP) acquired on different flakes. For most of the investigated zones (two of which are here given and labelled zones 1 and 2) the flake size is consistent with the SEM observations. In addition, the corresponding EFEDP evidences the disorganization of the sheets with diffraction spots arranged over rings characteristic of in-plane flake rotation with respect to each other. This is in contrast with zone 3, obtained in the same sample, where the flake is much larger and the EFEDP evidences well defined spots characteristic of the hexagonal symmetry (this kind of stack is largely in minority in this sample). This zone 3 is similar to those generally encountered in the $Ti_3C_2$-Li-24/25 sample, prepared with soft conditions. In addition to the disorder evidenced in the EFEDP, the C-K edge recorded in zones 1 and 2 shows a modification of the fine structure compared to the less perturbed zone 3. As evidenced in Fig. 6b, the A peak intensity is reduced, B peak is broadened and structures C and D are less pronounced in zones 1 and 2 compared to the C-K edge in zone 3, which is closer to the C-K edges recorded in samples obtained under softer etching conditions [60]. The broadening and intensity-loss observed in the fine structures spectra collected in zones 1 and 2, which are much more representative of most of the flakes than zone 3, is characteristic of structural disorder within the MXene sheets and evidences the effect of the harsh etching protocol on the structural quality of the flakes.

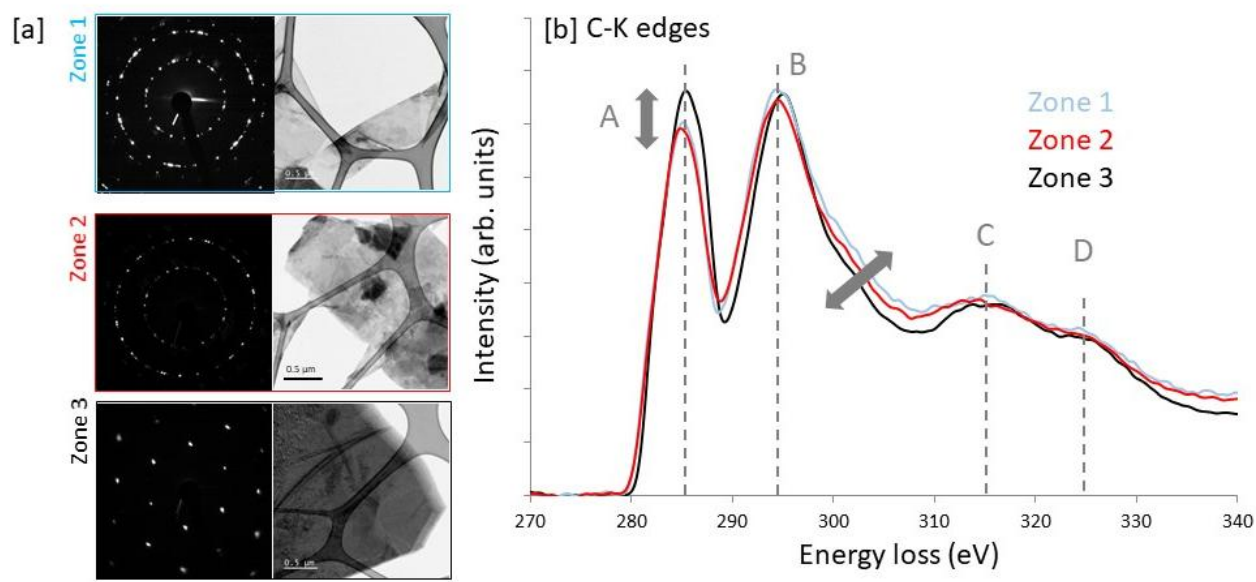

**Fig. 6.** [a] TEM micrographs and corresponding energy filtered electron diffraction patterns of different stacks in Ti$_3$C$_2$-Li-72/60 sample (*i.e.* prepared in harsh conditions). [b] C-K edges recorded on the three zones shown in (a).

*3.4 Summary of the different properties adjusted by the etching conditions*

Based on this study, it is clear that by varying the HF concentration in the etching agent, the surface properties as well as surface composition can be tuned. As shown by the XPS analyses performed on Ti$_3$C$_2$-HF10 and Ti$_3$C$_2$-HF48 samples, the amount of fluorine as terminal groups in the MXene can thus be controlled resulting in tunable ability to insert water between sheets as confirmed by XRD. Since -F terminal group is more hydrophobic than –OH and –O ones, this is consistent with the higher fluorine content in Ti$_3$C$_2$-HF48 that strongly limits the ability to intercalate water between sheets. This constitutes a major level in the optimization of MXenes for applications such as electrochemical devices, water treatment, catalysis or humidity sensors. Among the oxygen-based terminal groups, oxygen groups are favored when decreasing HF concentration. So, the variation of the conditions in HF medium allow tuning the O/OH ratio as terminal groups which can be interesting for specific applications preferentially involving one particular speciation.

Except for the presence of Li cations as well as a little amount of Cl in the LiF-etched MXenes, their composition is close to that of $Ti_3C_2$-HF10 sample, explaining why obtained Raman and XPS spectra are so similar. Nevertheless, the macrostructure of these three samples is quite different. Indeed, the Li-samples can be spontaneously dispersed and delaminated in water contrary to $Ti_3C_2$-HF10 [47]. This is certainly due to the presence of Li ions between the sheets whereas for $Ti_3C_2$-HF10 the washing step leads to a partial dehydration of the sample caused by the removal of cations inserted between the sheets during the etching step, thus disabling the delamination. As stated above, using the same initial LiF and HCl concentrations but different synthesis conditions - soft vs. harsh - results in the formation of MXenes with comparable compositions in general and similar terminal groups in particular (Table 3). However, when harsher etching conditions are used, they result in the formation of structural defects and smaller flakes. As discussed below, these defects greatly modify the surface properties of the samples.

All else being equal (temperature, duration, fluorine to MAX ratio), the $FeF_3$/HCl etching method leads to faster etching of Al compared to LiF/HCl etching. This method is thus of great interest when the efficiency of the etching process is the governing parameter in the synthesis of $Ti_3C_2T_x$ materials. Moreover Li is a limited natural resource[61], whereas iron is the fourth most abundant element in the earth's crust, which makes it more interesting from an industrial point of view. Furthermore, $FeF_3$/HCl etching results in $Ti_3C_2T_x@TiO_{2-x}F_{2x}$ composite in a one-pot synthesis. The MXene oxidation mechanism is closely related to the presence of Fe cations as Ti atoms in the MXene are oxidized by Fe cations during the synthesis. Indeed, if one assumes the average charge of Ti atoms in $Ti_3C_2T_x$ to be ≈ 2.4 [62], then the standard potential of the $Fe^{3+}/Fe^{2+}$ redox couple ($E°_{Fe3+/Fe2+}$ = 0.77 V vs. SHE) is higher than that of Ti ($E°_{Ti3+/Ti2+}$ = -0.37 V vs. SHE and $E°_{TiO(2+)/Ti3+}$ = +0.19 V vs. SHE). This oxidation probably favors the exfoliation step explaining why Al removal rate is higher compared to the samples synthesized using LiF/HCl as etching agent; despite similar synthesis conditions are used. Combining XPS and Raman results, it was shown that the oxidation was produced preferentially at the MXene sheet surfaces.

In our previous work, $FeF_3$/HCl was employed but at lower concentrations (initial F/Al atomic ratio equal to 7) [36] and no formation of $TiO_2$ (or $TiO_{2-x}F_{2x}$) was observed using Raman spectroscopy. To confirm this result, a $Ti_3C_2$-Fe-72/60 sample was also synthesized in this work with two times less $FeF_3$ in the etching solution (initial F/Al atomic ratio = 6) (See section V of ESI). In this case, a small amount of $TiO_2$ (or $TiO_{2-x}F_{2x}$) is observed in the Raman spectra (Fig. S6) compared to the other Fe-samples. Thus, by varying the synthesis parameters such as $FeF_3$ concentration, temperature and duration, it is possible to control in a one-pot synthesis the fraction of $TiO_2$, more accurately, an anatase-like $TiO_{2-x}F_{2x}$ structure. Taking into account that these composites are increasingly studied for different applications such as photocatalysis [63], sensors [13], batteries [64,65] or supercapacitors [66], our simple and one-pot synthesis allows for the facile control of the fraction of oxidized phase. On the other hand, by combining the positions of the XPS photopeaks and Raman bands, it seems that the formed oxidized phase is an oxyfluoride with anatase-like structure rather than an oxide.

*3.5 Investigation the electrochemical behavior of MXenes*

The voltammetric signatures of our MXenes were first investigated in a nitrogen-saturated 1 mol L$^{-1}$ KOH electrolyte at a scan rate of 50 mV s$^{-1}$. The resulting cyclic voltammograms for Ti$_3$C$_2$-HF10, Ti$_3$C$_2$-HF48, Ti$_3$C$_2$-Li-24/25 and Ti$_3$C$_2$-Li-72/60 materials are presented in Fig. 7a. For theses samples, an irreversible oxidation peak centered at *ca.* 0.9 – 1.1 V vs. RHE is observed during the positive scan of the first voltammetric cycle. In the absence of redox active species in the electrolyte, this redox peak can be most assuredly assigned to the irreversible oxidation of Ti atoms in direct contact with the electrolyte [67] and disappears in the following cycles as showed for Ti$_3$C$_2$- HF48 sample (Figure S8). This electrochemical response was found not to depend on the synthesis conditions since this intense oxidation peak was observed for all MXenes except for the Fe-samples (Fig. 7b and c). Indeed, the intensity of the irreversible oxidation peak for Ti$_3$C$_2$-Fe-24/25 (Fig. 7b) is much lower than those observed for the other confirming once again that this material was partly oxidized prior to voltammetric measurements. This decrease in intensity is even higher for Ti$_3$C$_2$-Fe-72/60 material (Fig. 7c), that is to know for the most oxidized sample, showing that this methodology is well-adapted to characterize the MXene surface oxidation. It is worth noting that the cyclic voltammograms obtained with FeF$_3$-etched samples have the typical I(E) behavior of materials possessing low electronic conductivities. The high oxide content of these samples probably strongly reduces their conductivity.

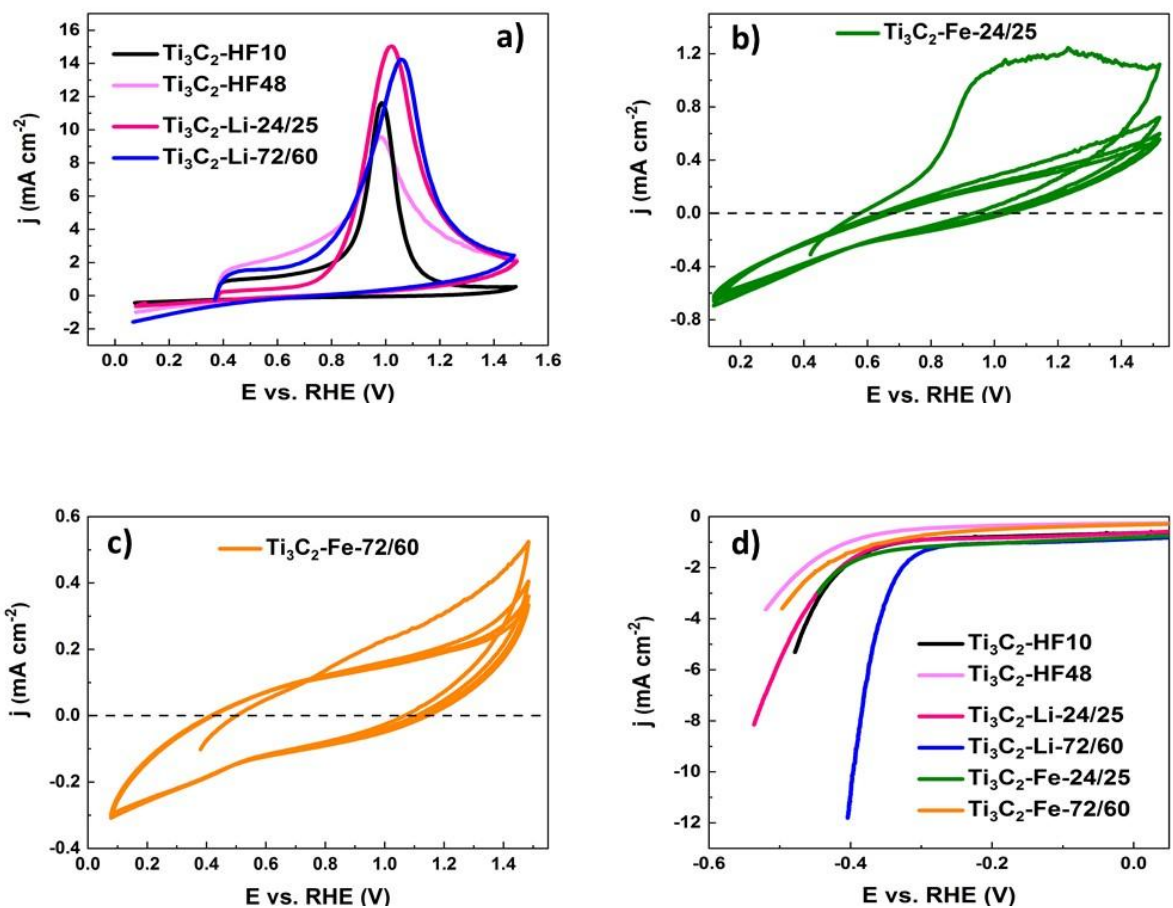

**Fig. 7.** Cyclic voltammograms of a) $Ti_3C_2$-HF10, $Ti_3C_2$-HF48, $Ti_3C_2$-Li-24/25, $Ti_3C_2$-Li-72/60 (only the first cycle is reported here for clarity), b) $Ti_3C_2$-Fe-24/25 and c) $Ti_3C_2$-Fe-72/60 samples recorded at a scan rate of 50 mV s$^{-1}$. d) HER polarization curves recorded with the different MXenes at a scan rate of 5 mV s$^{-1}$. All curves are recorded in a $N_2$ saturated 1 mol L$^{-1}$ KOH electrolyte.

The activity of the different MXenes towards HER was investigated by recording a polarization curve at a scan rate of 5 mV s$^{-1}$. HER is here used as a probe to shed light on the surface chemical stated of the MXenes. Fig. 7d summarizes our results and the obtained HER kinetic parameters are listed in Table 4. As observed, among the different studied MXenes, $Ti_3C_2$-Li-72/60 is the best catalyst for this reaction since both the overpotential (344 mV) and Tafel slope (96 mV dec$^{-1}$) are the lowest (Table 4). Focusing on the HF-etched samples, the HER activity of Ti$_3$C$_2$-HF10 is higher than the Ti$_3$C$_2$-HF48 material, the worst catalyst. These results are in agreement with the work of Handoko *et al*. [6] suggesting that the HER activity of MXene surfaces can be inhibited by the presence of terminal F atoms. The HER activity indeed decreases with the increase of F/Ti atomic ratio (Table 3). As discussed below, the F content is not systematically the main parameter which affects HER activity.

Focusing now on the Li-samples, Ti$_3$C$_2$-Li-72/60 MXene is largely better as an electrocatalyst for HER than Ti$_3$C$_2$-Li-24/25 (Table 4). Taking into account the similarity between the chemical compositions for these Li-containing samples (Table 3), this huge difference concerning HER activity cannot be solely related to the F content. In general, the experimental determination of the nature of the involved active site for HER is quite difficult. Nevertheless, the formation of smaller flakes (more exposed edges and corners,…) and more structural defects (as showed by TEM/EELS analyses) with Ti$_3$C$_2$-Li-72/60 sample change the nature of the active sites compared to Ti$_3$C$_2$-Li-24/25 explaining the difference in HER activities. As an example, Natu et al. showed that the charges of MXenes edges and surfaces are different [55], which can potentially drive to difference in the –OH adsorption kinetics for HER reactions.

HER activities of the Fe-samples (Fig. 6) are also relatively low in comparison with Ti$_3$C$_2$-Li-72/60 material. This is associated with both the surface oxidation and the low electronic conductivity. This is confirmed by the very low Tafel slopes of these samples compared to other MXenes. The lower activity of Ti$_3$C$_2$-Fe-72/60 compared to Ti$_3$C$_2$-Fe-24/25 can be ascribed to a higher fraction of oxidized Ti atoms. The role of oxidation on HER activity is further discussed below.

Focusing now on gravimetric capacitance (see ESI part VI for methodology), it is clear that Ti$_3$C$_2$-Li-24/25 sample is the most interesting among all materials with a gravimetric capacitance of 180 F g$^{-1}$ (table 4) indicating that MXene with larger flakes exhibit a higher capacitance as already reported by Peng *et al.* [68]. The high degree of oxidation of the FeF$_3$-etched samples precluded our measuring their capacitances.

To summarize, the synthesis conditions have a crucial role on electrochemical performances. As an example, the synthesis conditions (soft and harsh) applied using the LiF/HCl etching agent have a dramatic impact on surface properties of the resulting MXene though the chemical compositions are very similar. Soft synthesis condition favors a higher gravimetric capacitance by producing larger flakes containing a lower defect density. These materials can be used for applications such as supercapacitors, whereas harsh conditions favor the production of MXenes that are more active for the HER. The choice of the synthesis method is thus crucial and greatly depends on the targeted application.

**Table 4.** Capacitance and HER kinetic parameters obtained with different MXenes studies herein.

| MXene | Capacitance (F g$^{-1}$) | Overpotential $\eta_{j=3}$ (mV) at 3 mA cm$^{-2}$ | Tafel slope (mV dec$^{-1}$) |
|---|---|---|---|
| Ti$_3$C$_2$-HF10 | 110 | 440 | 165 |
| Ti$_3$C$_2$-HF48 | 100 | 500 | 195 |
| Ti$_3$C$_2$-Li-24/25 | 180 | 445 | 180 |
| Ti$_3$C$_2$-Li-72/60 | 130 | 344 | 96 |
| Ti$_3$C$_2$-Fe-24/25 | nd$^a$ | 445 | 200 |
| Ti$_3$C$_2$-Fe-72/60 | nd$^a$ | 480 | 285 |

$^a$ nd =not determined

*3.6 Correlation oxidation/HER/XPS*

To confirm the role of surface oxidation on HER performances, the HER activity of $Ti_3C_2$-HF48 was also measured after the irreversible surface oxidation occurring at around 1 V *vs.* RHE (potential cycling from -0.6 to 1.35 V *vs.* RHE) and compared with HER obtained after potential cycling from -0.6 to 0.5 V *vs.* RHE (Fig. 8), that not result in surface oxidation. As observed, the irreversible oxidation process leads to a significant decrease in the activity of MXene towards HER. Thus, oxidized MXene surfaces are not favorable for the HER.

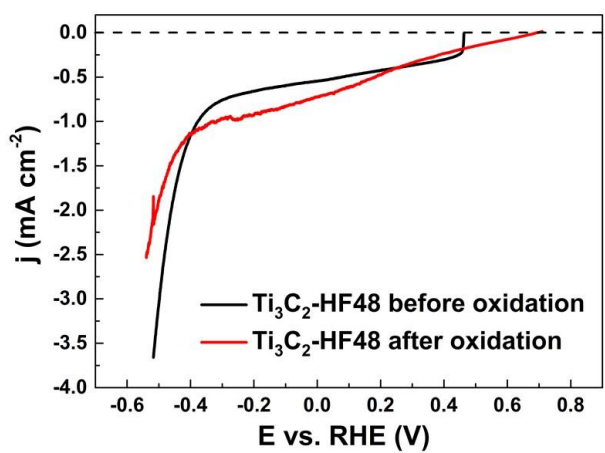

**Fig. 8.** HER polarization curves recorded with $Ti_3C_2$-HF48 before and after irreversible oxidation at a scan rate of 5 mV s$^{-1}$. All curves are recorded in a $N_2$ saturated 1 mol L$^{-1}$ KOH electrolyte.

Previous work showed that MXene surfaces are oxidation sensitive particularly under a wet atmosphere [36,69]. This behavior was used to confirm the role of Ti oxidation on HER activity. To further explore this aspect we studied the influence of storage conditions on MXene surface oxidation for the $Ti_3C_2$-HF10 and $Ti_3C_2$-HF48 samples. To do so, the fractions of oxidized Ti atoms of the fresh samples were compared with those of samples stored for one month either in a glove box under dry nitrogen (samples labelled as $Ti_3C_2$-HF10-$N_2$ and $Ti_3C_2$-HF48-$N_2$), or under ambient air (labelled as $Ti_3C_2$-HF10-air and $Ti_3C_2$-HF48-air).

Interestingly and irrespective of the storage conditions, the XRD pattern as well as the F contents - as determined by XPS - are similar to those of the initial samples. This is not surprising since MXenes powders are known to be relatively stable at a macroscopic scale as compared with MXene dispersed into a colloidal aqueous suspensions [70,71]. Nevertheless, the fraction of oxidized Ti atoms can be partly affected by the storage conditions. The high resolution XPS spectra of the Ti 2p region for the different samples are compared in Fig. 9. As observed, the fraction of oxidized Ti atoms at the surface is nearly the same for both fresh samples ($\approx$ 10%). After storage under $N_2$ and air atmospheres, the oxidation rate increases for $Ti_3C_2$-HF10, particularly under air (18.7%) whereas this rate does not evolve significantly for $Ti_3C_2$-HF48 samples. Thus, it can be deduced that the higher the F-content, the more stable the material is vis-a-vis oxidation after a month of storage.

The better stability of $Ti_3C_2$-HF10 material under a dry $N_2$ atmosphere in comparison with air is most probably related to the air humidity confirming already reported studies [36,69]. These different results are correlated with HER activity measurements. Indeed, as can be seen in Fig. 9, HER activity of the different $Ti_3C_2$-HF48 samples remains nearly constant, whereas the activity of the $Ti_3C_2$-HF10 samples decrease as the fraction of oxidized Ti atoms increases.

This result confirms that HER activity is intimately related to the surface oxidation state of samples and can thus be used as an accurate probe to characterize the state of oxidation of $Ti_3C_2$ surfaces. The decrease in the HER activity as the surface oxidation state is more pronounced is most likely related both to a decrease in the electronic conductivity of the MXene and to a chemical modification of the active sites for HER, leading to an increase in the charge transfer resistances.

Interestingly, the oxidation of the MXene surface can be partly avoided by an increase of the F-content for one month of storage duration. This is probably associated to the fact that this MXene is the only one of this work unable to insert water between MXene sheets, as evidenced by previously discussed XRD results. The few water inserted between sheets of interstratified $Ti_3C_2$-HF10 material (Fig. 1a), i.e. the main species responsible for MXene oxidation [69], can lead to a partial surface oxidation even during storage under a dry $N_2$ atmosphere.

Thus, to use MXenes as catalysts or active support for applications involving HER, controlling the oxidation kinetics is required. The balance between activity, higher for MXene with low F-contents and stability - which is better when water is not inserted between the sheets – is thus an important consideration that has to generally be taken into account.

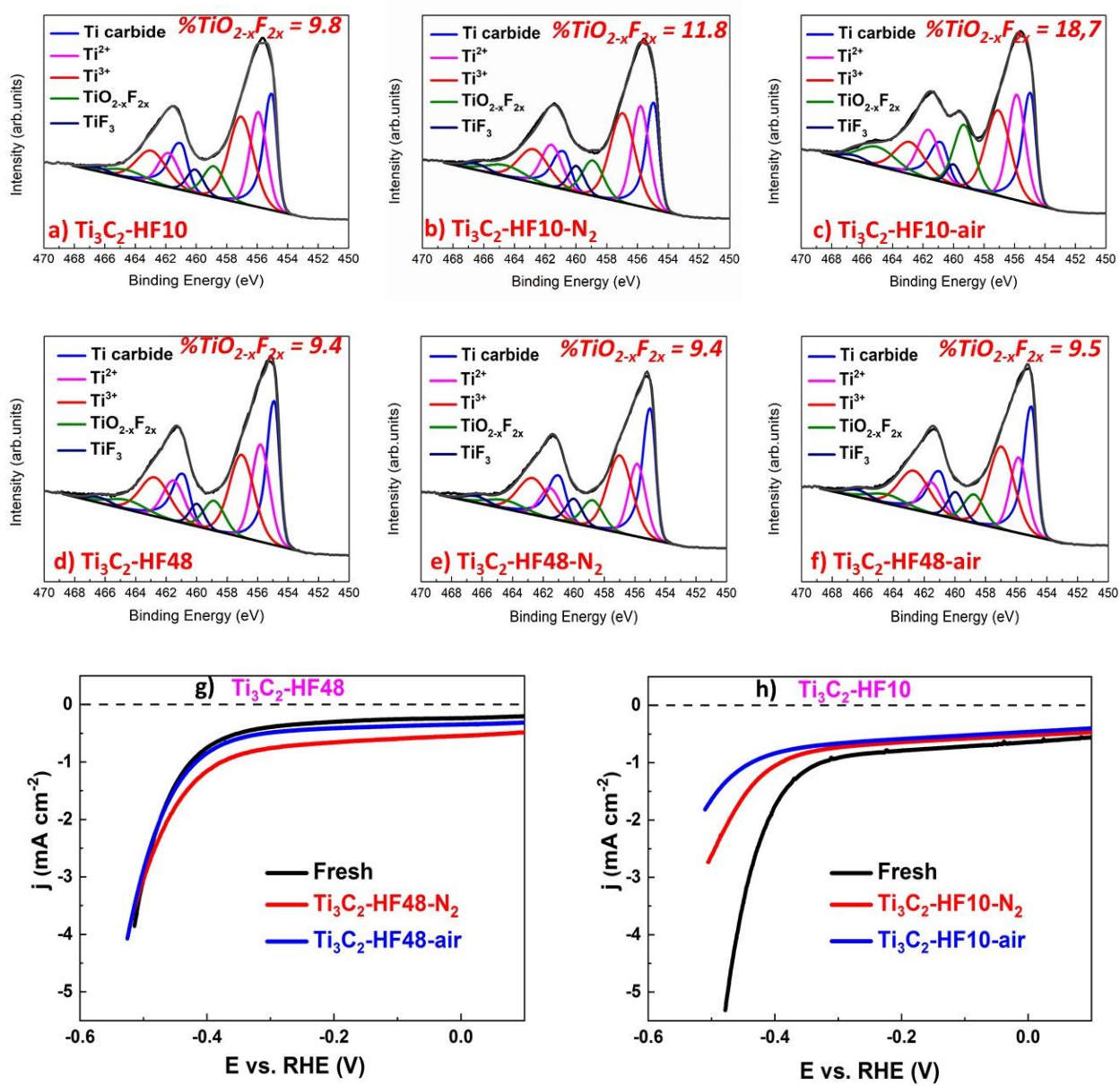

**Fig. 9.** XPS high resolution spectra of Ti 2p regions for a) fresh $Ti_3C_2$-HF10, and b) $Ti_3C_2$-HF48 and after aging under, c) and d) $N_2$ and e) and f) air atmospheres. g) $Ti_3C_2$-HF10 HER polarization curves and, h) $Ti_3C_2$-HF48 HER polarization curves after aging under different conditions at a scan rate of 5 mV s$^{-1}$ in a $N_2$ saturated 1 mol L$^{-1}$ KOH electrolyte.

## 4. Conclusions

In this study, the role of the etching agent on the modification of surface chemistries and properties of $Ti_3C_2T_x$ was explored. Three different etching agents were investigated: HF, LiF/HCl and $FeF_3$/HCl, each of them under both soft and harsh conditions. We showed that these etching agents lead to different surface chemistries and some properties.

Varying the HF concentration allows for varying the F content and consequently the ability to insert water. It was show that a low HF concentration allows the insertion of water layers probably stabilized by $H_3O^+$. Nevertheless, the washing on this sample, a crucial step for MXene synthesis, involves an irreversible decrease of the inserted water amount. The HF concentration also allows modifying the distribution of terminal groups (O/OH ratio) as well as the stability against surface oxidation. Each of these surface chemistries is interesting for various applications. For example, at lower HF concentrations, the formation of MXenes with a higher activity is favoured probably due to its lower F content. At higher HF concentrations the stability against oxidation is favoured for at least one month.

Etching with LiF/HCl results in producing conductive clays facilitating the processing of these materials. Soft etching conditions - low temperatures and durations - led to the formation of $Ti_3C_2T_x$ with large flakes and high capacitances. If the etching conditions are harsh (high temperature and duration) the resulting flakes are small and more defective leading to an enhanced HER activity.

When the LiF is replaced by $FeF_3$, the kinetics of the MAX to MXene transformation is enhanced, even when the initial F/Al ratio is the same, thanks to the oxidation properties of $Fe^{3+}$, which is interesting to reduce the global cost of the MXene production. On the other hand, this method do not allow a spontaneous delamination in water as in the case of LiF/HCl etching method.

This method allows also the formation of $Ti_3C_2T_x@TiO_{2-x}F_{2x}$ powders. The tuning of the synthesis parameters (temperature, duration, Fe concentration) allows for the control of the amount of $TiO_{2-x}F_{2x}$ formed during synthesis. This oxyfluoride compound have the anatase structure. Producing this kind of composite with controlled composition in a one-pot synthesis is interesting for numerous applications such as photocatalysis and batteries.

By contrast, this work highlights that MXene surface oxidation should be avoided for HER application.

Thus, the synthesis parameters, even using a same etching agent solution, can have a significant impact on the surface properties. For all the above-mentioned reasons, this work could serve as a guideline for the MXene community to help in choosing the appropriate etching agent to obtain the desired physicochemical properties required for any targeted application.

Undoubtedly, the exploration of these wonderful new materials and the future improvements in synthesis methods are still in their infancy and other remarkable properties may be discovered, further expending the range of potential applications.

Another major aim of this work was to improve our understanding of XRD, Raman and XPS signals obtained on $Ti_3C_2T_x$. This was possible thanks to the sensitive tuning in surface composition, ability to insert water, oxidation, over-etching, amorphous carbon formation and macrostructural properties provided by the study of different etching agents. This study validates different theoretical and experimental studies on XPS and Raman spectroscopy provided in the literature to characterize these materials possessing a high degree of chemical complexity. Beyond the confirmation of the composition evolution based on XPS analyses, we show in this study that Raman spectroscopy is a simple and powerful tool to characterize the MXene terminal groups. Finally, we showed that HER can be also an accurate surface probe to characterize surface properties of MXenes in addition to conventional methods. In brief, we believe that this work can also serve as a guideline for the MXene community of suitable characterization tools for an in-depth understanding of MXenes surface chemistry.

5. Acknowledgments

The authors acknowledge financial support from the "Agence National de la Recherche" (reference ANR-18-CE08-014 – MXENECAT project), the European Union (ERDF), the "Région Nouvelle Aquitaine" and the French research ministry (Ph.D. thesis of M. Benchakar). This work also partially pertains to the French Government program "Investissements d'Avenir" (LABEX INTERACTIFS, reference ANR-11-LABX-0017-01) which is here gratefully acknowledged for the financial support to C. Garnero post-doctoral position. The authors wish to thank J. Rousseau for her assistance on SEM analyses.

6. Associated content

**Electronic Supporting Information ESI (separate file)**

Description of the $Ti_3AlC_2$ MAX phase synthesis with the corresponding XRD pattern (Fig. S1); XRD patterns of $Ti_3C_2$-HF10 and $Ti_3C_2$-HF48 before and after washing steps (Fig. S2); XRD patterns of the slurry and sediment of $Ti_3C_2$-Li-24/25 (Fig. S3); global atomic percentages (Table S1) and XPS peak fitting results (Table S2 to S7) for the different studied MXenes; XPS results for commercial $TiF_3$ (Fig. S4); SEM analyses of the different MXenes (Fig. S5); EDS characterization and specific surface area of the different MXenes (Table S8); Description of Ti$_3$C$_2$-Fe-72/60 − n$_{FeF3}$/2 synthesis and Raman spectra of the FeF$_3$-etched samples (Fig. S6); description of the strategy to determine capacitance (Fig. S7). Voltammetric cycles for Ti$_3$C$_2$-HF48 (Fig. S8).